\documentclass[10pt, onecolumn, conference]{IEEEtran}

\usepackage{enumitem}
\usepackage{amsmath} 
\usepackage{graphicx,url,tabularx,booktabs}
\usepackage[english,brazil]{babel}
\usepackage[utf8]{inputenc}  
\usepackage{comment}
\usepackage{ifthen}
\usepackage{graphicx}
\usepackage{tcolorbox}
\usepackage{amssymb}
\usepackage{pifont}
\usepackage[table]{xcolor}
\usepackage{multirow}

\usepackage{float}
\usepackage{breakcites}
\usepackage{makecell}
\usepackage{tikz}
\usetikzlibrary{arrows.meta,positioning,fit,backgrounds}
\title{AnonShield: Scalable On-Premise\\Pseudonymization for CSIRT Vulnerability Data}

\author{
\IEEEauthorblockN{
    Cristhian Kapelinski\IEEEauthorrefmark{1},
    Douglas Lautert\IEEEauthorrefmark{1},
    Beatriz Machado\IEEEauthorrefmark{1},\\
    Diego Kreutz\IEEEauthorrefmark{1},
    Isadora Garcia Ferrão\IEEEauthorrefmark{2}
}
\IEEEauthorblockA{\IEEEauthorrefmark{1}Federal University of Pampa (UNIPAMPA)}
\IEEEauthorblockA{\IEEEauthorrefmark{2}Université de Bretagne Occidentale (UBO)}
}
\begin{document} 
\selectlanguage{english}
\maketitle
     
\begin{abstract}
We present AnonShield, a high-throughput, on-premise pseudonymization system that combines GPU-accelerated NER, streaming processing, caching, and schema-aware configuration. Evaluated on datasets up to 550\,MB (70{,}951 records), AnonShield reduces processing time from over 92 hours to under 10 minutes (up to 738$\times$ speedup) while achieving up to 94.2\% F1-score and 96.7\% recall.
Our results show that scalable pseudonymization of vulnerability data is feasible without sacrificing analytical utility, enabling compliant data sharing in operational CSIRT environments.
\end{abstract}


\section{Introduction}
\label{sec:intro}

Computer Security Incident Response Teams (CSIRTs) routinely process vulnerability scan reports generated by tools such as OpenVAS and Tenable across large and heterogeneous infrastructures, including containerized services, legacy systems, campus and backbone networks. These reports embed sensitive information, such as Personally Identifiable Information (PII), IP addresses, hostnames, and credentials, alongside technical findings~\cite{Kapelinski2025}. Sharing such data across organizations or machine learning pipelines exposes these identifiers to potential adversaries, raising compliance challenges under GDPR, LGPD, and internal data governance policies. Indeed, network identifiers like IP addresses may qualify as personal data under GDPR when linked to individuals~\cite{Albakri2019,Nweke2020}. As vulnerability datasets grow in volume, addressing this exposure at operational scale is critical for compliant data sharing and for enabling downstream uses such as LLM-based analysis and the generation of synthetic cyber threat data~\cite{almorjan2025synthetic}

The scale of this problem is rapidly increasing. In 2025, the National Vulnerability Database recorded 48{,}448 CVEs, a 20\% increase over 2024, and forecasts indicate approximately 59{,}400 CVEs in 2026~\cite{cvedetails2025,FIRST2026}. VulnCheck reported 884 Known Exploited Vulnerabilities in 2025, with nearly 29\% exploited at or before disclosure~\cite{VulnCheck2026}. In Brazil, CAIS/RNP continuously scans its infrastructure using Tenable, generating datasets exceeding 70{,}000 vulnerability records per cycle (Dataset~D2). Such volume makes manual analysis unfeasible and demands LLM-assisted processing. However, feeding raw operational data into external pipelines conflicts with data sovereignty and GDPR/LGPD policies. Regulatory guidance under LGPD recognizes pseudonymization as a key mechanism for enabling lawful data processing~\cite{ANPD2024}, ensuring that downstream tools, such as LLM-based analyzers, receive context-rich but identifier-free records~\cite{Bandel2025}. Vulnerability scan data primarily raises security and sovereignty concerns; incident data additionally exposes PII of victims, adding a direct LGPD/GDPR compliance dimension. The tension between sharing vulnerability data and protecting embedded PII has been documented in Cyber Threat Intelligence literature~\cite{Albakri2019,Nweke2020,Wagner2016}, yet no widely adopted solution exists for large-scale vulnerability scan reports.

Table~\ref{tab:related_work} positions existing approaches across five dimensions relevant to CSIRT deployments. The solution space divides into three categories, each with structural limitations. Commercial cloud frameworks such as Google Cloud DLP~\cite{GoogleDLP} and Amazon Comprehend~\cite{AmazonComprehend} provide scalable pipelines with extensive entity coverage but require data transmission to external endpoints, violating data sovereignty constraints. Microsoft Presidio~\cite{MicrosoftPresidio}, while on-premise, lacks recognizers for cybersecurity-specific entities such as CVE identifiers, CPE strings, certificate serial numbers, and cryptographic artifacts, leaving parts of the vulnerability profile exposed. IRI DarkShield~\cite{DarkShield} exhibits similar limitations and introduces licensing restrictions that hinder adoption in academic and public-sector CSIRTs. Open-source tools address narrower problems. Anonip~\cite{Anonip} supports only IP anonymization, ARX~\cite{Prasser2014} focuses on tabular data and may degrade analytical utility through $k$-anonymity~\cite{Slijep2021}, and LogLicker~\cite{LogLicker2023} relies on static RegEx patterns that lead to higher false-negative rates in heterogeneous vulnerability data.

\begin{table}[!htp]
\centering
\caption{{\scriptsize Overview of anonymisation and pseudonymisation
tools and their primary gap relative to the CSIRT vulnerability
processing context. Data type: 
\textbf{US}~=~Unstructured; \textbf{S}~=~Structured.
\textbf{--} in Formats: tool operates on in-memory strings with no
native file I/O.}}
\label{tab:related_work}
\renewcommand{\arraystretch}{1.3}
\resizebox{\textwidth}{!}{%
\scriptsize
\begin{tabular}{|
    >{\raggedright\arraybackslash}p{2.2cm}|
    >{\raggedright\arraybackslash}p{3.4cm}|
    >{\raggedright\arraybackslash}p{1.9cm}|
    >{\raggedright\arraybackslash}p{0.7cm}|
    >{\raggedright\arraybackslash}p{2.8cm}|
    >{\raggedright\arraybackslash}p{4.2cm}|}
\hline
\textbf{System} &
\textbf{Main Technique} &
\textbf{Target Domain} &
\textbf{Type} &
\textbf{File Formats} &
\textbf{Gap in CSIRT context} \\
\hline

Microsoft Presidio~\cite{MicrosoftPresidio} &
NLP + RegEx + extensible recognizers &
Generic text & US &
-- &
No CVE/CPE/hash recognizers; domain-irrelevant patterns produce FPs
in vulnerability data \\
\hline

Google Cloud DLP~\cite{GoogleDLP} &
NLP + FPE + PII classification &
General data & US+S &
\texttt{.txt, .csv, .json,} \texttt{.pdf, .docx} &
Cloud endpoint conflicts with CSIRT data sovereignty policies;
no cybersecurity-specific entities \\
\hline

Amazon Comprehend~\cite{AmazonComprehend} &
Cloud NLP for PII detection and redaction &
Generic text & US &
\texttt{.txt} &
Cloud endpoint conflicts with CSIRT data sovereignty policies;
no vulnerability scanner format support \\
\hline

IRI DarkShield~\cite{DarkShield} &
Multi-source masking + REST API &
Corporate data & US+S &
\texttt{.txt, .pdf, .xml,} \texttt{.json, .sql, images} &
Commercial licence limits adoption in academic and public-sector
CSIRTs; no cybersecurity-specific entities \\
\hline

LogLicker~\cite{LogLicker2023} &
RegEx + masking + remapping manifest &
CloudTrail \& Generic logs & US+S &
\texttt{.json, .txt} &
RegEx only; lacks contextual NER; high manual maintenance for custom logs \\
\hline
 
Anonip~\cite{Anonip} &
Bit masking of trailing IP bits via pipe &
Web server logs & US &
\texttt{.log, .txt} &
Covers IP addresses only \\
\hline
 
ARX~\cite{Prasser2014} &
$k$-anonymity, $l, t, \delta$ privacy models &
Biomedical data & S &
\texttt{.csv, .xlsx, sql} &
Tabular data only; inapplicable to vulnerability reports \\
\hline
 
AnonLFI v1.0~\cite{Bandel2025} &
Hybrid NER + RegEx + SHA-256 (no salt) &
Incidents & US+S &
\texttt{.txt, .docx,} \texttt{.csv, .xlsx} &
Saltless SHA-256 vulnerable to rainbow-table attacks; no native XML/JSON support; impractical at scale \\
\hline

AnonLFI v2.0~\cite{Kapelinski2025} &
Hybrid NER + RegEx + OCR + HMAC-SHA256 + native XML/JSON processors &
Incidents \& Vulnerabilities & US+S &
\texttt{.xml, .json, .pdf,} \texttt{.txt, .csv, .docx,} \texttt{.xlsx,} images &
Fails at operational scale; superlinear latency from DOM parsing; no GPU or streaming; high variability (CV $>$ 0.96); $>$92 hours for 550\,MB \\
\hline

\textbf{AnonShield (this work)} &
\textbf{Hybrid NER + RegEx + OCR + HMAC-SHA256 + GPU + LRU cache + streaming} &
\textbf{Incidents \& Vulnerabilities} & \textbf{US+S} &
\texttt{.xml, .json, .pdf,} \texttt{.txt, .csv, .docx,} \texttt{.xlsx, .jsonl, .log,} images &
  \\
\hline
 
\end{tabular}}
\end{table}

Prior AnonLFI generations were the first solutions tailored to CSIRT contexts. AnonLFI v1.0~\cite{Bandel2025} introduced a hybrid NER and RegEx pipeline validated on 763 incidents, achieving Precision of 100\% and Recall of 97.38\%, but relies on saltless SHA-256 and lacks native support for XML and JSON. AnonLFI v2.0~\cite{Kapelinski2025} extended support to vulnerability data, adding HMAC-SHA256, native XML and JSON processing, and OCR capabilities, achieving F1-scores of 76.5\% for OCR and 92.13\% for OpenVAS XML. However, its estimated processing time exceeds 92 hours for operational datasets with 70{,}951 records, making it impractical at scale. Taken together, no existing solution simultaneously satisfies four key requirements: cryptographic pseudonymization with referential integrity, native processing of hierarchical formats, domain-specific entity recognition, and scalable on-premise execution.

This paper presents AnonShield, a  pseudonymization framework that addresses the limitations of prior frameworks and tools. AnonShield extends prior designs with GPU-accelerated NER, LRU caching, streaming I/O, and a schema-aware \texttt{anonymization\_config} mechanism that enables per-field policies without NER overhead. The contributions are threefold: (1) a scalable pseudonymization architecture for CSIRT vulnerability data, (2) a comparative evaluation of four strategies (\texttt{standalone}, \texttt{hybrid}, \texttt{presidio}, \texttt{filtered}) on datasets of up to 550\,MB (70{,}951 records), including comparisons with AnonLFI v1.0 and v2.0, and (3) an accuracy evaluation using a statistically representative sample of 67 vulnerability records annotated by three specialists. The evaluation spans both a controlled testbed with 130 services and an operational dataset from RNP infrastructure, demonstrating the feasibility of large-scale, on-premise pseudonymization.

\section{The AnonShield Framework}
\label{sec:anonshield}
AnonShield\footnote{\label{fn:anonshield_repo}\url{https://github.com/AnonShield/tool}} is a pseudonymisation framework for CSIRTs processing vulnerability scan reports and incident data. It detects and replaces sensitive entities with deterministic pseudonyms, preserving referential integrity across correlated reports. All processing is performed locally, mitigating privacy risks associated with third-party data sharing and supporting GDPR~\cite{Amoo2024} and LGPD requirements.

\begin{figure}[!htp]
    \centering
    \includegraphics[width=1\linewidth]{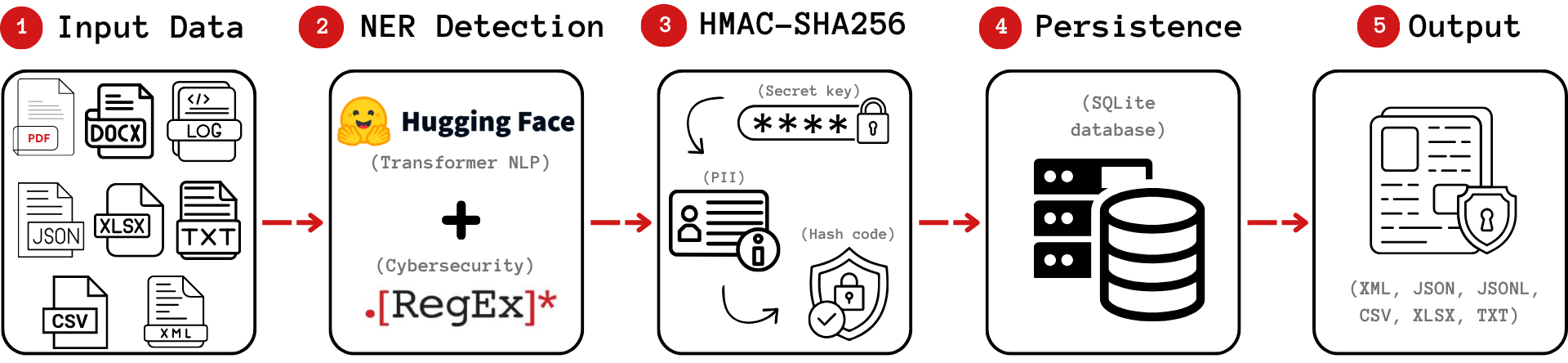}
   \caption{AnonShield pipeline. The same entity produces a consistent pseudonym across documents given the same secret key.}
    \label{fig:pipeline}
\end{figure}

The framework is organised into five sequential stages, illustrated in Figure~\ref{fig:pipeline}.

\textbf{(1)~Input.}
AnonShield supports XML, JSON, JSONL, CSV, TXT, LOG, PDF (text and image), DOCX, and XLSX formats. Large files are processed via incremental \textit{streaming}, using \texttt{ijson} for JSON and SAX parsing for XML, avoiding full in-memory loading.

\textbf{(2)~Entity Detection (NER).}
Detection combines a configurable transformer model, \texttt{Davlan/xlm-roberta-base-ner-hrl} (default, multilingual) or \texttt{attack-vector/SecureModernBERT-NER} (cybersecurity-focused), with 21 custom RegEx patterns covering IPv4/IPv6, URLs, hostnames, hashes, certificates, CVE/CPE identifiers, credentials, and others. Inference runs on GPU (CUDA), with an LRU cache to amortise repeated lookups across records.

\textbf{(3)~HMAC-SHA256 Pseudonymisation.}
Each entity is replaced by an HMAC-SHA256 slug computed with an operator-defined secret key. The slug length is configurable from 0 to 64 hexadecimal characters (up to 256\,bits of entropy), and the entity type is preserved as a prefix, e.g., \texttt{100.111.20.23}~$\to$~\texttt{[IP\_ADDRESS\_48624b5cdc]}.

\textbf{(4)~Persistence.}
The entity~$\leftrightarrow$~slug mapping is stored in a local SQLite database (\texttt{entity\_type}, \texttt{original\_name}, \texttt{slug\_name}, \texttt{full\_hash}, \texttt{first\_seen}, \texttt{last\_seen}), enabling controlled re-identification by an administrator holding both the database and the HMAC key. The database supports persistent and in-memory modes.

\textbf{(5)~Output.}
For structured formats (XML, JSON, JSONL, CSV, XLSX), the original schema and structure are preserved, with only sensitive values replaced. For PDF, DOCX, and image inputs, text is extracted via parsing or OCR, pseudonymised, and written to plain-text (\texttt{.txt}) output files.

AnonShield recognises 34 entity types across six categories, extending beyond classical PII to include cybersecurity-specific identifiers such as CVE IDs, CPE strings, UUIDs, and OIDs. Although publicly available, these identifiers may expose an organization’s vulnerability profile and enable targeted exploitation, making their sensitivity context-dependent. To address this, AnonShield decouples entity detection from replacement policy, enabling fine-grained control via \texttt{--entities-to-preserve} and \texttt{fields\_to\_exclude}, which support selective pseudonymization and field-level exclusion without altering the recognition pipeline.

AnonShield supports multiple anonymisation strategies, selectable via \texttt{-{}-anonymization-strategy}, to balance accuracy and performance. Presidio maximises recall but introduces false positives from domain-irrelevant recognizers. Filtered mitigates this by restricting recognizers to cybersecurity-relevant entities, while Hybrid preserves detection behaviour and reduces overhead through a lightweight replacement mechanismy, yielding identical accuracy. Standalone bypasses Presidio entirely, achieving the highest throughput and lowest latency at a small recall cost due to span-boundary differences. An experimental SLM-based strategy, exploring locally deployed small language models for entity detection, is currently under development and remains outside the scope of this paper's evaluation. Additionally, the schema-aware \texttt{anonymization\_config} enables per-field control through \texttt{force\_anonymize}, \texttt{fields\_to\_anonymize}, and \texttt{fields\_to\_exclude}, improving both accuracy and performance and supporting reusable templates for formats such as STIX~2.x. Full details, including entity coverage, strategies, and configuration options, are available in the tool's GitHub repository\textsuperscript{\ref{fn:anonshield_repo}}.

\section{Experimental Methodology}
\label{sec:method}

The evaluation uses four datasets (Table~\ref{tab:datasets_summary}), covering
different scanning engines, operational scales, and processing challenges.

\begin{table}[!htp]
\centering
\caption{Dataset summary.}
\label{tab:datasets_summary}
\resizebox{0.8\textwidth}{!}{
\renewcommand{\arraystretch}{1.3}
\begin{tabular}{lllrl}
\toprule
\textbf{ID} & \textbf{Source} & \textbf{Format} & \textbf{Size} & \textbf{Access} \\
\midrule
\textbf{D1}  & OpenVAS (VulnLab) & CSV/TXT/PDF/XML & 9.29/11.61/33.32/34.11\,MB & Public \\
\textbf{D1C} & Converted from D1 & XLSX/DOCX/JSON/PDF(img) & 3.34/7.58/60.72/1{,}479.56\,MB & Public \\
\textbf{D2}  & Tenable (CAIS/RNP)& CSV/JSON & 419.72/550.54\,MB & Private \\
\textbf{D3}  & Synthetic (Mock)  & CSV/JSON & 247.45/444.56\,MB & Public \\
\bottomrule
\end{tabular}
}
\end{table}

\textbf{D1 -- OpenVAS Heterogeneous Dataset.} 
We scanned 130 targets from the VulnLab testbed across eight categories (Table~\ref{tab:vulnlab_categories}), yielding 520 reports in four formats. These reports, available in the AnonShield repository\textsuperscript{\ref{fn:anonshield_repo}}, include OpenVAS native \texttt{.anonymous} XML files, which demonstrate limitations in built-in redaction: while hostnames are masked, IP addresses, environment names, and internal UUIDs are retained in the output. Furthermore, 11 reports contained 71 unique cryptographic hashes and TLS fingerprints. Since these identifiers are often indexable by search engines, they may facilitate asset re-identification, indicating that native anonymization is insufficient for preventing information disclosure in these contexts.

\begin{table}[!htp]
\centering
\caption{VulnLab service categories.}
\label{tab:vulnlab_categories}
\renewcommand{\arraystretch}{1.3}
\begin{tabular}{ll}
\hline
\textbf{Category} & \textbf{Examples} \\
\hline
Network \& Infrastructure  & DNS (Bind), FTP (ProFTPD), OpenLDAP \\
Web Applications \& APIs   & Juice Shop, DVWA, GraphQL \\
Web Servers \& Platforms   & Apache, Nginx, Tomcat, WordPress \\
Databases                  & MySQL, PostgreSQL, MongoDB, Redis \\
Monitoring \& Logging      & Elasticsearch, Prometheus, Grafana \\
DevOps \& CI/CD            & Jenkins, GitLab, Nexus \\
Messaging \& Streaming     & Kafka, RabbitMQ, Zookeeper \\
Operating Systems          & Ubuntu (14/16), Debian (8/9), CentOS (6/7) \\
\hline
\end{tabular}
\end{table}

\textbf{D1C -- Converted Formats.}
Each D1 report was converted to XLSX, DOCX, JSON, and image-only PDF, yielding
520 additional files. Rasterized PDFs inflate mean size by ${\approx}44\times$,
forcing the pipeline to rely exclusively on its OCR subsystem.

\textbf{D2 -- CAIS/RNP Operational Dataset.}
Restricted Tenable scans comprising 70{,}951 vulnerability records (JSON 550\,MB;
CSV 420\,MB), representative of large-scale enterprise environments and used as the
primary operational benchmark. Not publicly released due to privacy and security constraints. In fact, publishing vulnerability scan data is infeasible even after partial PII removal: retaining a vulnerability name or identifier drastically reduces an attacker's search space. Removing all such identifiers to achieve safe publication would strip the dataset of its core content: a vulnerability dataset without vulnerability data has no analytical value. Consequently, for this data type, the most viable path to external sharing is synthetic data generation (an approach increasingly adopted for threat intelligence via LLMs \cite{almorjan2025synthetic}), while pseudonymization remains essential for enabling internal use of operational data in LLM pipelines without compromising organizational sovereignty or violating GDPR/LGPD. Pseudonymization and synthetic generation are thus complementary: anonymized data feeds LLMs that produce shareable synthetic datasets.

\textbf{D3 -- Synthetic Dataset.} 
A public digital twin of D2, generated to replicate its structural and statistical characteristics. Unique identifiers from the original dataset were mapped to random entries from the official CVE list\footnote{\url{cve.org/downloads}}. This process ensures that D3 maintains the same frequency distribution and entity density as the operational dataset while eliminating sensitive organizational information and specific vulnerability findings.

\textbf{Hardware.}
All experiments ran on a single workstation: NVIDIA RTX 5060 Ti (16\,GB VRAM),
AMD Ryzen 5 8600G (6c/12t), 32\,GB DDR5-6000 RAM.

\textbf{Accuracy Evaluation.}
A statistically representative sample ($n{=}67$, population 6{,}472; 90\% CL,
$E{=}10\%$) was annotated by three domain specialists across 13 entity types,
yielding TP, FP, and FN counts for Precision, Recall, and F1-Score. AnonShield
used \texttt{SecureModernBERT-NER}; a preservation list (\texttt{TOOL},
\texttt{MALWARE}, \texttt{CVSS}) and \texttt{force\_anonymize} policies for
pattern-less fields (e.g., \texttt{Hostname}) were applied uniformly across all
versions under comparison.

\textbf{Statistical Analysis.}
A four-stage protocol was applied: Shapiro-Wilk normality test ($\alpha{=}0.05$);
Mann-Whitney U with Benjamini-Hochberg correction ($p_{\text{adj}}{<}0.05$);
Cohen's $d$ for effect size; and power-law regression ($T{=}a{\cdot}S^{\alpha}$)
combined with polynomial fitting to separate fixed overhead from throughput scaling.

\section{Results}
\label{sec:results}

This section presents a comprehensive empirical evaluation of AnonShield across two complementary dimensions: processing performance and pseudonymization accuracy. The analysis spans heterogeneous and converted file formats, large-scale operational datasets, and a specialist-annotated validation set, enabling a rigorous assessment of scalability, throughput, statistical significance, and privacy-preserving effectiveness under realistic CSIRT workloads. 

\subsection{Processing Performance}

\textbf{Heterogeneous formats (D1).}
Table~\ref{tab:d1_scientific_performance} (Appendix~\ref{sec:results_d1})
reports latency across 520 files ($N{=}260$ runs/strategy).
AnonShield\_standalone consistently achieves the lowest latency and highest
stability across all formats, with speedups of $16.51\times$ (XML),
$9.56\times$ (CSV), $3.27\times$ (PDF), and $3.04\times$ (TXT) over
AnonLFI~v2.0, all statistically meaningful sizes (Cohen's $d \geq 0.42$).
AnonLFI~v2.0 exhibits pathological variability ($CV > 0.96$) caused by
memory-intensive DOM parsing; AnonShield reduces this to $CV < 0.8$ through
iterative streaming and GPU inference. Notably, AnonLFI~v2.0 regresses against
AnonLFI~v1.0 on CSV and TXT due to increased recognizer complexity without
architectural compensation, a limitation addressed by AnonShield.
Scalability analysis (Figure~\ref{fig:d1_scalability}, Appendix~\ref{sec:results_d1})
confirms near-linear complexity ($\alpha \approx 1$) for AnonShield, with an
amortization effect driving throughput above 60\,KB/s for larger files, while
AnonLFI~v2.0 degrades superlinearly.

\textbf{Converted formats (D1C).}
Results for XLSX, DOCX, JSON, and image-only PDF are detailed in
Table~\ref{tab:converted_perf} (Appendix~\ref{sec:results_d1c}).
AnonShield\_standalone achieves a $23.24\times$ speedup on JSON ($CV{=}0.60$),
$5.00\times$ on XLSX, and $2.08\times$ on DOCX. For image-only PDF, where the
${\approx}44\times$ size inflation shifts the bottleneck to OCR, speedup narrows
to $1.64\times$ while maintaining linear scaling ($R^2 \geq 0.989$). A single
file (\textit{openssh-server\_images.pdf}) caused 2 AnonShield failures due to
a malformed content stream triggering \texttt{PyMuPDF}'s strict parser, an
edge case absent in AnonLFI~v2.0 only because it lacks the corresponding
memory-management step.

\textbf{Operational scale (D2 and D3).}
Table~\ref{tab:large_scale_d2_merged} and Table~\ref{tab:large_scale_d3_merged}
(Appendix~\ref{sec:results_large_scale}) present the most consequential results
of this evaluation. On D2 (70{,}951 records, 550\,MB JSON), AnonShield\_standalone
completes processing in $453\,\text{s}$ ($1{,}250\,\text{KB/s}$), whereas
AnonLFI~v2.0's estimated runtime exceeds ${\sim}$92 hours, a speedup of
${\geq}738\times$. Activating the schema-aware \texttt{anonymization\_config}
reduces D2-CSV runtime further to $12.55\,\text{s}$ ($34{,}341\,\text{KB/s}$),
a $46.9\times$ gain over full NER inference, by enabling deterministic
field-level remapping that bypasses the inference pipeline entirely.
On D3, GPU acceleration yields speedups of up to $3{,}532\times$ over
AnonLFI~v2.0; even on CPU-only hardware, AnonShield\_standalone achieves
${\geq}535\times$ speedup, confirming that performance gains are architectural
rather than hardware-dependent. Throughput on D3 ($3{,}473\,\text{KB/s}$, CSV)
exceeds D2 ($732\,\text{KB/s}$) due to higher entity redundancy enabling
effective LRU cache reuse; D2's lower cache hit rate, driven by high-entropy
operational data, reflects realistic worst-case deployment conditions.
All strategy comparisons are statistically significant ($p_{\text{adj}} < 0.001$,
Mann-Whitney U with Benjamini-Hochberg correction).

\subsection{Accuracy}

Table~\ref{tab:accuracy} presents results on the 67-record specialist-annotated
validation set (13 entity types). Recall is the operationally critical metric:
an unredacted False Negative exposes sensitive infrastructure details to adversaries and constitutes a potential GDPR/LGPD breach, whereas a False Positive degrades analytical utility without compromising security or compliance
Accuracy improves substantially across generations: AnonLFI~v1.0 achieves
F1~$=$~23.7\% (Recall~$=$~18.8\%); AnonLFI~v2.0 reaches 54.0\%
(Recall~$=$~40.7\%); and AnonShield \texttt{filtered}/\texttt{hybrid} achieve
F1~$=$~94.2\% at Recall~$=$~96.7\%. The \texttt{presidio} strategy matches this
Recall but at lower Precision (71.6\%), favoring high-sensitivity environments
where over-anonymization is acceptable. \texttt{Standalone} trails by 2.2\,pp
in Recall (94.5\%) at the highest throughput, a gap attributable to
span-boundary behavior rather than a fundamental model limitation.

All 25 common False Negatives across AnonShield strategies originate from two
sources: (1)~19 unrecognized organization names
(\textit{Oracle}, \textit{ISC}, \textit{Wiesemann \& Theis}) in formulaic
vulnerability attribution strings, where entity context is insufficient for
NER disambiguation; and (2)~6 structural artifacts in semi-structured
free-text, including port values appearing as plain integers
(e.g., \texttt{80 or 443}), where neither regex nor contextual NER can infer
entity type without schema information. These failure modes are deterministic
and enumerable, making them addressable via targeted recognizer extensions.
False Positives are dominated by Presidio's \texttt{DATETIME} and
\texttt{IP\_ADDRESS} recognizers misclassifying version strings
(e.g., \texttt{2.4.51}~$\to$~\texttt{[DATE\_TIME\_...]}); \texttt{filtered}
and \texttt{hybrid} suppress this to 64 FPs via a curated recognizer subset.
A systematic AnonLFI~v2.0 misclassification of \texttt{Severity:~Medium} as
\texttt{ORGANIZATION} is fully resolved in AnonShield via
\texttt{anonymization\_config}. Full False Negative and False Positive
breakdowns are provided in Appendix~\ref{sec:accuracy_results}.

\section{Availability, Implementation and Demonstration Plan}
\label{sec:availability}

AnonShield is publicly available on GitHub\textsuperscript{\ref{fn:anonshield_repo}}, including source code, documentation, datasets, and all evaluation artefacts. The repository provides step-by-step installation instructions, a minimal functional test executable in under 5 minutes, and scripts to reproduce the main experimental results. The tool is implemented in Python (3.12), uses Microsoft Presidio for NER orchestration, and is managed via \texttt{uv}, with no external service dependencies. It is also distributed as a Docker image in CPU and GPU (CUDA) variants on Docker Hub\footnote{\url{https://hub.docker.com/r/anonshield/anon}}.

For the SBRC Tools Lounge, the demonstration will run on a standard laptop (4 vCPUs, 8 GB RAM) running Ubuntu, without requiring specialized hardware or network infrastructure. A GPU is optional, as the CPU version is fully functional. The demonstration includes live pseudonymization of OpenVAS reports, comparison of anonymization strategies, and illustration of throughput gains enabled by the schema-aware \texttt{anonymization\_config} mechanism.

\section{Conclusion and Future Work}
\label{sec:conclusion}

This paper presented AnonShield, the third generation of the AnonLFI research line, achieving up to ${\sim}738\times$ speedup (from ${\sim}$92.9 hours to under 10 minutes on 550\,MB) and ${\geq}535\times$ on CPU-only hardware. The \texttt{filtered} and \texttt{hybrid} strategies deliver the best accuracy-performance trade-off (F1~=~94.2\%, Recall~=~96.7\%), while \texttt{anonymization\_config} provides additional gains of up to $47\times$. These results demonstrate that large-scale, on-premise pseudonymization is operationally practical for CSIRT workflows.

Two boundaries define the current scope. First, the framework is constrained by evaluation and processing limits: the accuracy baseline relies on a restricted sample size and dataset diversity, format extraction suffers from OCR/PyMuPDF breaks, and algorithms face NER context starvation and \texttt{standalone} truncation. To structurally bypass format issues, we are extracting vulnerabilities directly as structured datasets from PDFs~\cite{machado2025}. Second, formal privacy models primarily suit incident data with victim PII; for vulnerability contexts, publishing raw records remains inherently infeasible regardless of anonymization depth.

As future work, we plan to advance AnonShield in four directions: improving robustness for complex formats, especially image-only PDFs and malformed streams; extending entity recognition with cybersecurity-specific recognizers and locally deployed SLMs to reduce false negatives; expanding the evaluation to larger and more diverse CSIRT datasets; and investigating formal privacy-utility trade-off models for secure data sharing and LLM-assisted cyber analytics.

\bibliographystyle{plain}
\bibliography{sbc-template}

\appendix

\section{Performance and Scaling Analysis (Dataset D1)}
\label{sec:results_d1}

\begin{figure}[!htp]
\centering
\includegraphics[width=0.9\textwidth]{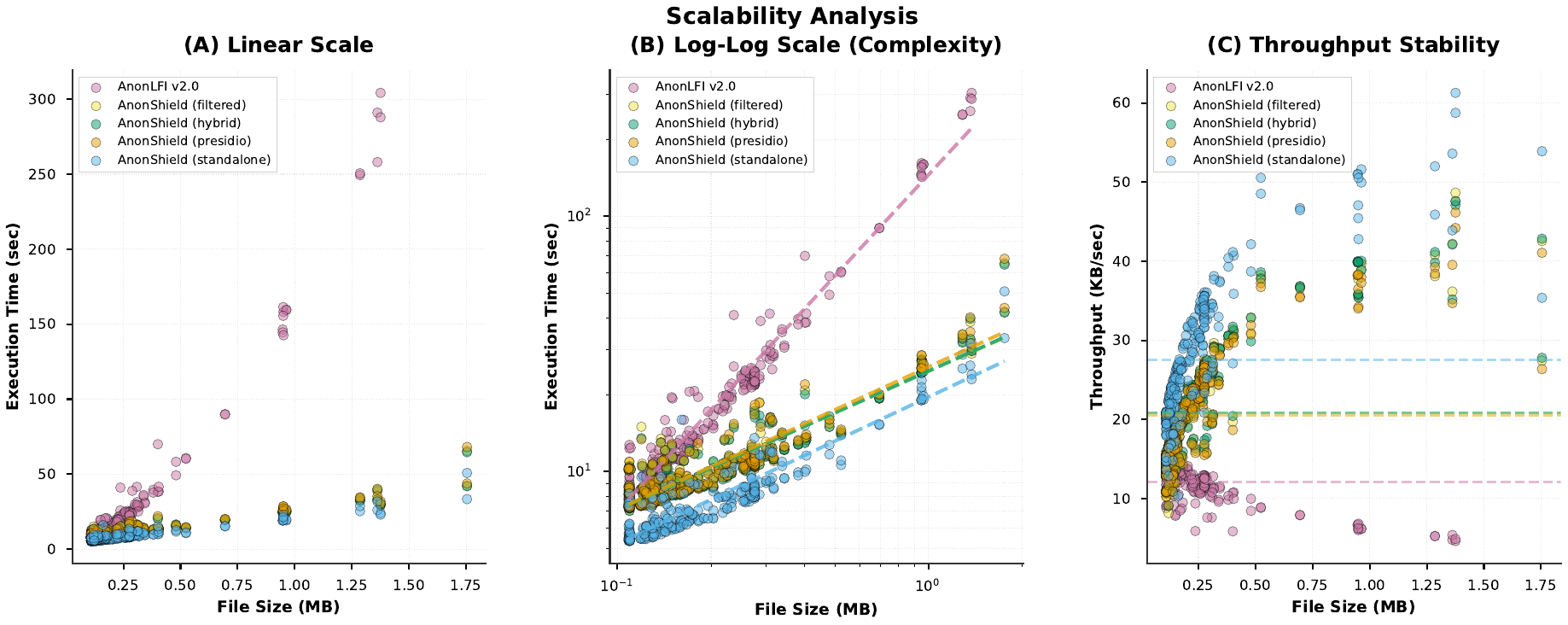}
\caption{Scalability analysis on D1 PDF subset: (A) Linear scale, (B) Log-Log complexity ($\alpha \approx 1$ for AnonShield), and (C) Throughput stability.}
\label{fig:d1_scalability}
\end{figure}

Performance metrics for 520 files ($N=260$ runs/strategy) are summarized in Table~\ref{tab:d1_scientific_performance}. Due to file-size heterogeneity and right-skewed distributions (Shapiro-Wilk $p < 10^{-6}$), the median is used as a robust measure for D1, while the mean ($\bar{t} \pm \sigma$) is reserved for D2/D3. Across these runs, the \texttt{standalone} strategy consistently outperforms Presidio-based versions ($1.3\times$--$1.4\times$ faster) by bypassing orchestration overhead. Further scalability insights (Figure~\ref{fig:d1_scalability}, Panel C) highlight an amortization effect: as file size increases, AnonShield\_standalone throughput rises (peaking $>60$~KB/s), effectively spreading fixed startup costs, whereas AnonLFI~v2.0 shows degrading throughput, confirming its inefficiency for operational-scale workloads.

\begin{table}[!htp]
\centering
\caption{Processing latency on D1 reports.}
\label{tab:d1_scientific_performance}
\renewcommand{\arraystretch}{1.3}

\begin{tabular}{llrrrrrrc}
\toprule
\textbf{Format} & \textbf{Version/Strategy} & \textbf{Mean (s)} & \textbf{Median (s)} & \textbf{Max (s)} & \textbf{CV} & \textbf{Speedup} & \textbf{Cohen's $d$} \\
\midrule
\multirow{3}{*}{\textbf{XML}} & AnonLFI~v2.0 (Baseline) & 191.89 & 176.44 & 1717.71 & 0.96 & $1.00\times$ & -- \\
 & AnonShield\_presidio & 15.62 & 13.69 & 85.58 & 0.60 & $12.28\times$ & 1.29 (Large) \\
 & \textbf{AnonShield\_standalone} & \textbf{11.62} & \textbf{10.02} & \textbf{72.34} & \textbf{0.63} & $\mathbf{16.51\times}$ & \textbf{1.32 (Large)} \\
\midrule
\multirow{4}{*}{\textbf{CSV}} & AnonLFI~v2.0 (Baseline) & 74.18 & 37.01 & 1442.08 & 1.75 & $1.00\times$ & -- \\
 & AnonLFI~v1.0 & 37.92 & 20.41 & 577.96 & 1.55 & $1.96\times$ & 0.35 (Small) \\
 & AnonShield\_presidio & 10.65 & 9.08 & 60.62 & 0.53 & $6.97\times$ & 0.67 (Medium) \\
 & \textbf{AnonShield\_standalone} & \textbf{7.76} & \textbf{6.67} & \textbf{47.87} & \textbf{0.53} & $\mathbf{9.56\times}$ & \textbf{0.70 (Medium)} \\
\midrule
\multirow{3}{*}{\textbf{PDF}} & AnonLFI~v2.0$^{\dag}$ (Baseline) & 27.38 & 13.01 & 304.24 & 1.70 & $1.00\times$ & -- \\
 & AnonShield\_presidio & 11.20 & 9.25 & 68.13 & 0.58 & $2.44\times$ & 0.47 (Small) \\
 & \textbf{AnonShield\_standalone} & \textbf{8.36} & \textbf{6.91} & \textbf{50.78} & \textbf{0.60} & $\mathbf{3.27\times}$ & \textbf{0.56 (Medium)} \\
\midrule
\multirow{4}{*}{\textbf{TXT}} & AnonLFI~v2.0 (Baseline) & 31.23 & 12.61 & 747.74 & 2.20 & $1.00\times$ & -- \\
 & AnonLFI~v1.0 & 20.28 & 10.98 & 354.91 & 1.56 & $1.54\times$ & 0.19 (Negligible) \\
 & AnonShield\_presidio & 13.10 & 10.21 & 100.28 & 0.73 & $2.38\times$ & 0.36 (Small) \\
 & \textbf{AnonShield\_standalone} & \textbf{10.28} & \textbf{7.91} & \textbf{80.71} & \textbf{0.78} & $\mathbf{3.04\times}$ & \textbf{0.42 (Small)} \\
\bottomrule
\end{tabular}

{\scriptsize $^{\dag}$~AnonLFI~v2.0 PDF failed runs excluded due to out-of-memory errors. $^{\dag}$$^{\dag}$~Presidio-based strategies are statistically similar ($p_{adj} > 0.46$), with AnonShield\_presidio as representative. AnonLFI~v1.0 does not support XML or PDF.}
\end{table}

\section{Converted Format Benchmarks (D1C)}
\label{sec:results_d1c}

To evaluate versatility, 130 reports were converted into XLSX, DOCX, JSON, and image-only PDF ($N=260$ runs/strategy). Table~\ref{tab:converted_perf} summarizes the results, with significant differences across strategies ($p < 10^{-5}$). AnonLFI~v2.0 regressed on XLSX and DOCX due to additional recognizers without architectural optimizations, while the 2 failed AnonShield runs on image-only PDFs were caused by \texttt{PyMuPDF} parser errors triggered by malformed content streams.

\begin{table}[!htp]
\centering
\caption{Processing latency on D1-converted formats.}
\label{tab:converted_perf}
\renewcommand{\arraystretch}{1.3}

\begin{tabular}{llrrrrrr}
\toprule
\textbf{Format} & \textbf{Version/Strategy} & \textbf{Mean (s)} & \textbf{Max (s)} & \textbf{CV} & \textbf{Speedup} & \textbf{Cohen's $d$} \\
\midrule
\multirow{3}{*}{\textbf{JSON}}
 & AnonLFI~v2.0 (Baseline) & 246.55 & 1861.01 & 1.16 & $1.00\times$ & -- \\
 & AnonShield\_presidio            &  20.15 &   110.78 & 0.72 & $12.23\times$ & 1.08 (Large) \\
 & \textbf{AnonShield\_standalone}&  \textbf{10.61} & \textbf{49.02} & \textbf{0.60} & $\mathbf{23.24\times}$ & \textbf{1.12 (Large)} \\
\midrule
\multirow{4}{*}{\textbf{XLSX}}
 & AnonLFI~v1.0 (Baseline) & 35.51 & 320.73 & 1.37 & $1.00\times$ & -- \\
 & AnonLFI~v2.0            & 60.30 & 596.95 & 1.52 & $0.59\times$ & $-0.34$ (Small) \\
 & AnonShield\_presidio            &  9.86 &  38.64 & 0.47 & $3.60\times$ & 0.73 (Medium) \\
 & \textbf{AnonShield\_standalone}&  \textbf{7.11} & \textbf{27.95} & \textbf{0.47} & $\mathbf{5.00\times}$ & \textbf{0.82 (Large)} \\
\midrule
\multirow{4}{*}{\textbf{DOCX}}
 & AnonLFI~v1.0 (Baseline) & 19.58 & 186.55 & 1.34 & $1.00\times$ & -- \\
 & AnonLFI~v2.0            & 29.90 & 431.60 & 1.94 & $0.65\times$ & $-0.23$ (Small) \\
 & AnonShield\_presidio            & 12.16 &  63.24 & 0.68 & $1.61\times$ & 0.36 (Small) \\
 & \textbf{AnonShield\_standalone}&  \textbf{9.43} & \textbf{51.51} & \textbf{0.73} & $\mathbf{2.08\times}$ & \textbf{0.52 (Medium)} \\
\midrule
\multirow{3}{*}{\textbf{PDF-image$^{\dag}$}}
 & AnonLFI~v2.0 (Baseline) & 59.15 & 757.63 & 1.85 & $1.00\times$ & -- \\
 & AnonShield\_presidio            & 37.85 & 362.81 & 1.46 & $1.56\times$ & 0.23 (Small) \\
 & \textbf{AnonShield\_standalone}& \textbf{36.04} & \textbf{348.63} & \textbf{1.54} & $\mathbf{1.64\times}$ & \textbf{0.27 (Small)} \\
\bottomrule
\end{tabular}

{\scriptsize
$^{\dag}$ Failures: 1 file (\textit{openssh-server\_images.pdf}) failed in both runs ($N=2$) across all AnonShield strategies. $^{\dag}$$^{\dag}$\texttt{AnonShield\_presidio} represents the Presidio cluster. Speedup for JSON/PDF-image vs. AnonLFI~v2.0; XLSX/DOCX vs. AnonLFI~v1.0.}
\end{table}

\section{Large-Scale Operational Benchmarks (D2 and D3)}
\label{sec:results_large_scale}

AnonShield was evaluated against operational datasets D2 (Tenable, 550\,MB) and D3 (Mock CVE, 444\,MB). Due to the prohibitive slowness of AnonLFI v1.0 and v2.0, their runtimes were estimated based on D1 throughput. For CSV, these are \textbf{conservative lower bounds} (${\geq}121$ hours), as v2.0’s superlinear scaling significantly inflates costs as file sizes grow. For D3, an additional CPU-only benchmark (no GPU) was conducted to assess whether AnonShield's throughput gains depend on GPU hardware.

\begin{table}[!htp]
\centering
\caption{D2 (Operational) processing times (70,951 records).}
\label{tab:large_scale_d2_merged}
\renewcommand{\arraystretch}{1.3}

\begin{tabular}{llrrrr}
\toprule
 & & \multicolumn{2}{c}{\textbf{CSV (419.72\,MB)}} & \multicolumn{2}{c}{\textbf{JSON (550.54\,MB)}} \\
\cmidrule(lr){3-4}\cmidrule(lr){5-6}
\textbf{Mode} & \textbf{Version / Strategy} & \textbf{Time (s)} & \textbf{KB/s} & \textbf{Time (s)} & \textbf{KB/s} \\
\midrule
\multirow{3}{*}{\textbf{No config}}
 & AnonLFI~v2.0 \emph{(est.)} & ${\geq}437{,}335$ & 0.98 & ${\sim}334{,}472$ & 1.69 \\
 & AnonShield\_presidio & $2{,}520.7 \pm 68.0$ & 171 & $985.7 \pm 73.7$ & 575 \\
 & \textbf{AnonShield\_standalone} & $\mathbf{588.5 \pm 30.7}$ & \textbf{732} & $\mathbf{453.1 \pm 35.9}$ & \textbf{1{,}250} \\
\midrule
\multirow{2}{*}{\textbf{With config}}
 & AnonShield\_presidio & $13.42 \pm 0.13$ & 32{,}034 & $18.88 \pm 0.16$ & 29{,}855 \\
 & \textbf{AnonShield\_standalone} & $\mathbf{12.55 \pm 0.74}$ & \textbf{34{,}341} & $\mathbf{18.03 \pm 0.23}$ & \textbf{31{,}272} \\
\midrule
\multicolumn{2}{l}{\textbf{Speedup standalone vs. v2.0 (no config)}} & \multicolumn{2}{c}{$\mathbf{{\geq}743\times}$} & \multicolumn{2}{c}{$\mathbf{{\sim}738\times}$} \\
\bottomrule
\end{tabular}

\scriptsize AnonShield\_standalone is significantly faster than Presidio-based strategies ($p_{adj} < 0.001$).
\end{table}

\textbf{Analysis of Large-Scale Impact.}
Without schema-aware configuration (full NER inference active), the GPU variant is up to ${\sim}6.6\times$ faster than CPU on CSV and ${\sim}5.1\times$ on JSON; with config (inference bypass), both perform similarly (${\sim}8\text{--}9$\,s).

\begin{table}[!htp]
\centering
\caption{D3 (Synthetic) processing times.}
\label{tab:large_scale_d3_merged}
\renewcommand{\arraystretch}{1.3}

\begin{tabular}{llrrrr}
\toprule
 & & \multicolumn{2}{c}{\textbf{CSV (247.45\,MB)}} & \multicolumn{2}{c}{\textbf{JSON (444.56\,MB)}} \\
\cmidrule(lr){3-4}\cmidrule(lr){5-6}
\textbf{Mode} & \textbf{Version / Strategy} & \textbf{Time (s)} & \textbf{KB/s} & \textbf{Time (s)} & \textbf{KB/s} \\
\midrule
\multirow{3}{*}{\textbf{No config (GPU)}}
 & AnonLFI~v2.0 \emph{(est.)} & ${\geq}257{,}835$ & 0.98 & ${\sim}270{,}086$ & 1.69 \\
 & AnonShield\_presidio & $318.0 \pm 5.1$ & 797 & $339.6 \pm 14.4$ & 1{,}343 \\
 & \textbf{AnonShield\_standalone} & $\mathbf{73.0 \pm 1.6}$ & \textbf{3{,}473} & $\mathbf{172.1 \pm 6.2}$ & \textbf{2{,}647} \\
\midrule
\textbf{No config (CPU)} & \textbf{AnonShield\_standalone} & $\mathbf{481.5 \pm 8.9}$ & \textbf{526} & $\mathbf{881.9 \pm 57.7}$ & \textbf{518} \\
\midrule
\multirow{2}{*}{\textbf{With config}}
 & AnonShield\_presidio & $8.89 \pm 0.06$ & 28{,}517 & $21.11 \pm 0.22$ & 21{,}571 \\
 & \textbf{AnonShield\_standalone} & $\mathbf{7.96 \pm 0.08}$ & \textbf{31{,}856} & $\mathbf{20.43 \pm 0.81}$ & \textbf{22{,}314} \\
\midrule
\multicolumn{2}{l}{\textbf{Speedup standalone (GPU) vs. v2.0}} & \multicolumn{2}{c}{$\mathbf{{\geq}3{,}532\times}$} & \multicolumn{2}{c}{$\mathbf{{\sim}1{,}569\times}$} \\
\multicolumn{2}{l}{\textbf{Speedup standalone (CPU) vs. v2.0}} & \multicolumn{2}{c}{$\mathbf{{\geq}535\times}$} & \multicolumn{2}{c}{$\mathbf{{\sim}306\times}$} \\
\bottomrule
\end{tabular}

{\scriptsize Standalone vs. Presidio comparisons are statistically significant in all cases ($p_{adj}<0.001$).}
\end{table}

\section{Accuracy Evaluation}
\label{sec:accuracy_results}

Partial anonymizations were counted as 1\,TP + 1\,FN. For \texttt{standalone}, all counts were computed against actual occurrences in the original text rather than the pseudonym count in the output: lacking an overlap-resolution layer, \texttt{standalone} can generate more pseudonyms than real occurrences (e.g., different recognizers detecting overlapping spans independently), which would artificially inflate TP and FP counts if the output were used as the counting basis. The \texttt{standalone} Recall gap (2.2\,pp) is a mechanical artifact of this same span-boundary behavior: its sequential replacement occasionally truncates entities (e.g., \texttt{[IP\_ADDRESS\_...].1.72}) or merges overlapping patterns.

\begin{table}[!htp]
\centering
\caption{Accuracy on 67-record validation set (D1 OpenVAS CSV).}
\label{tab:accuracy}
\renewcommand{\arraystretch}{1.3}

\resizebox{0.5\linewidth}{!}{
\begin{tabular}{lrrrrrrr}
\toprule
\textbf{Version/Strategy} & \textbf{TP} & \textbf{FP} & \textbf{FN}
  & \textbf{Prec.} & \textbf{Rec.} & \textbf{F1} \\
\midrule
AnonLFI v1.0               & 106 & 224 & 459 & 32.1\% & 18.8\% & 23.7\% \\
AnonLFI v2.0               & 279 &  68 & 407 & 80.4\% & 40.7\% & 54.0\% \\
\midrule
AnonShield\_presidio   & 724 & 287 & 25  & 71.6\% & 96.7\% & 82.3\% \\
AnonShield\_filtered   & 724 &  64 & 25  & 91.9\% & 96.7\% & 94.2\% \\
AnonShield\_hybrid     & 724 &  64 & 25  & 91.9\% & 96.7\% & 94.2\% \\
AnonShield\_standalone & 739 & 102 & 43  & 87.9\% & 94.5\% & 91.1\% \\
\bottomrule
\end{tabular}
}
\end{table}

\textbf{False Negatives.} All AnonShield strategies share 25 FNs from two sources: 19 unrecognized organization names due to NER context starvation in formulaic strings, and 6 structural artifacts in free-text that cannot be disambiguated without schema information. \texttt{Standalone} adds 18 FNs caused by span-boundary fragmentation, leading to partial information leaks. These results reinforce a critical principle: \textbf{anonymizing $k{-}1$ out of $k$ occurrences is equivalent to anonymizing none}, as a single leak enables exact re-identification.

\textbf{False Positives.} \texttt{AnonShield\_presidio} produces 287 FPs, mainly from misclassified version strings and domain-irrelevant patterns. \texttt{Filtered} and \texttt{hybrid} reduce this to 64 through curated recognizers, while \texttt{Standalone} reaches 102 FPs due to the absence of Presidio’s validation and overlap-resolution mechanisms. A systematic FP in AnonLFI~v2.0 is fully resolved via \texttt{anonymization\_config}.

\section{Research Artefacts}
\label{sec:artefacts}
All reproducibility artefacts and scripts are available at: \url{https://github.com/AnonShield/tool/blob/main/paper_data/AVAILABILITY.md}.

\end{document}